\documentclass[prl,twocolumn,a4paper,showpacs,superscriptaddress]{revtex4}

\usepackage{graphicx}
\usepackage{amsmath}

\begin{document}

\title{$\theta$ dependence of $\textrm{CP}^9$ model}

\date{May 26, 2003}
\author{V.~Azcoiti}
%\email{azcoiti@azcoiti.unizar.es}
\affiliation{Departamento de F\'{\i}sica Te\'orica, Universidad de
Zaragoza,
Cl. Pedro Cerbuna 12, E-50009 Zaragoza (Spain)}
\author{G.~Di Carlo}
%\email{gdicarlo@lngs.infn.it}
\affiliation{INFN, Laboratori Nazionali del Gran Sasso, 67010 Assergi, 
(L'Aquila) (Italy)}
\author{A.~Galante}
%\email{galante@lngs.infn.it}
\affiliation{INFN, Laboratori Nazionali del Gran Sasso, 67010 Assergi, 
(L'Aquila) (Italy)}
\affiliation{Dipartimento di Fisica dell'Universit\`a di L'Aquila,
67100 L'Aquila (Italy)}
\author{V.~Laliena}
%\email{laliena@posta.unizar.es}
\affiliation{Departamento de F\'{\i}sica Te\'orica, Universidad de
Zaragoza,
Cl. Pedro Cerbuna 12, E-50009 Zaragoza (Spain)}

\begin{abstract}
We apply to the $\textrm{CP}^9$ model 
two recently proposed numerical techniques for simulation of
systems with a $\theta$ term.
The algorithms, successfully tested in the strong coupling limit, 
are applied to the weak coupling region. 
The results agree and errors have been evaluated and are at \% level.
The results scale well with the renormalization group equation 
and show that, for $\textrm{CP}^9$ 
in presence of a $\theta$-term,
$\textrm{CP}$ symmetry is spontaneously broken at $\theta=\pi$
in the continuum limit.
\end{abstract}
\maketitle

Understanding the role of the $\theta$ parameter in QCD and its connection
with the ``strong $\textrm{CP}$ problem'' is one of the major challenges
for high energy theorists \cite{Strongcp}. 
Unfortunately euclidean Lattice Gauge Theory,
our main non perturbative tool for QCD studies, has not been able
to help us for two reasons: {\it i}) the difficulties in defining 
the topological charge operator on the lattice \cite{Top_def}
and {\it ii}) the
imaginary contribution to the action that comes from the $\theta$-term
prevents the applicability of the importance sampling method.

Concerning the $\theta$-term effect on the low energy dynamics,
two dimensional $\textrm{CP}^{N-1}$ models are very interesting since
they have $\theta$-vacua, a well defined topological charge 
and many features in common with four-dimensional Yang-Mills theories:
they are asymptotically free, possess instanton solutions, and 
admit a $1/N$ expansion.
Of course they share point {\it ii)} with QCD and this means that,
from the point of view of numerical calculations, the complex
action problem is as severe as in QCD.
Unfortunately the algorithms used in the past to tackle the complex
action problem are limited to moderate lattice sizes and
the fate of $\textrm{CP}$ symmetry at $\theta=\pi$ is still an open 
question \cite{Plefka,Japan_2}.

Recently we proposed two independent methods for simulating models with
$\theta$-term. Both are based on standard ({\it i.e.} with real action) 
simulations of the model at imaginary $\theta=-ih$ but then they use
completely different strategies. The first is based on the reconstruction
of the probability distribution function of the order parameter
\cite{metodo1}, the second introduces a particular transformation that 
allows to reconstruct the order parameter via an extrapolation of a 
suitably chosen quantity \cite{metodo2}.
Both can in principle suffer from systematics but the agreement of the 
two gives, a posteriori, a very stringent indication that 
systematic effects are indeed under control.

We tested the two algorithms in some models where the analytical solution
is available, and also a perfect agreement between the two approaches 
was found for $\textrm{CP}^{3}$ in the
strong coupling region. The aim of present work is to extend this study
to the continuum limit.

Moving to the weak coupling region is a tricky issue.
$\textrm{CP}^{N-1}$ models have the problem of unphysical
configurations characterized by topological structures of the size
of one lattice spacing (called dislocations) \cite{Dis_1,Dis_2}.
The effect is relevant for small
$N$ while, for $N\to \infty$, is suppressed and the
model becomes a system of $N$ free
particles and $N$ free antiparticles where $\theta$-vacua are degenerate 
in energy \cite{N_1,N_2}. 
If we are interested in the continuum limit the effect of dislocations
has to be negligible not to spoil the scaling properties of the theory.
For small $N$ this forces to go to 
very large values of the correlation length
and, consequently, untenable large volumes are mandatory. 

Previous studies of $\textrm{CP}^{N-1}$ models at $\theta=0$
reported that $N=4$ is not large enough to suppress dislocations
\cite{Campostrini}.
Indications of scaling of the topological susceptibility
$\chi_t(\theta=0)$ were observed but violations
of universality were also present: simulations with
different actions lead to distinct asymptotic values
\cite{Campostrini,burkhalter}. 
It is not the scope of present work to analyze the contribution
of dislocations to the topological charge at finite $\theta$,
and it is not easy to address
this problem within our framework since different definition
of the charge operator (like the one based on local polynomial 
definition or cooling) can not be directly applicable to 
$\theta=-ih$ simulations.
What we know is that increasing $N$ the situation is less 
severe and indeed, using the results of Campostrini et al \cite{Campostrini},
%after some preliminary study, 
we can affirm that $N=10$ is large enough. Therefore 
we concentrated our efforts on the scaling region of the 
$\textrm{CP}^9$ model.

We have adopted for
the action the ``auxiliary U(1) field'' formulation
\begin{equation}
S_g = N\beta\sum_{n,\mu}(\bar{z}_{n+\mu}z_n U_{n,\mu} +
\bar{z}_n z_{n+\mu} \bar{U}_{n,\mu}-2)
\label{action}
\end{equation}
where $z_n$ is a $N$-component complex scalar field that satisfies
$\bar{z}_n z_n = 1$ and $U_{n,\mu}$ is a $U(1)$ ``gauge field''.
The topological charge operator is defined directly from the $U(1)$ field:
\begin{equation}
S_\theta = 
i \frac{\theta}{2\pi}\sum_p\log(U_p) 
\label{topocharge}
\end{equation}
where $U_p$ is the product of the $U(1)$ field around the
plaquette and $-\pi < \log(U_p) \le \pi$.

For both methods the action to simulate $S=S_g + S_{\theta=-ih}$ is
real and local. This implies that the main computational cost
is practically equivalent to standard simulations
of $\textrm{CP}^{N-1}$ model at $\theta=0$. 
This is important since, as will be clear in the following,
large volume simulations 
were mandatory to have good quality raw data.

We refer to the original papers for more information and 
summarize here the main points of the new methods.

The first one \cite{metodo1,lat02_vic} reconstructs 
the p.d.f. of the topological charge density that is introduced 
to define the finite volume partition function as a discrete sum over the
different topological sectors:
\begin{equation}
{\cal Z}_V(\theta)\;=\;\sum_{x}\,e^{-V f_V(x)}\,
{\mathrm e}^{{\mathrm i}\theta V x} \, .
\label{Z}
\end{equation}
where $x=n/V$ is the density of topological charge ($n$ is an integer
that labels sectors with different topological charge) and
$e^{-Vf_V(x)}$ is, up to a normalization constant, 
the above mentioned p.d.f. .
From the measurement of the mean value of the density of topological charge 
as a function of $h$ we use the saddle point approximation to 
get, up to ${\cal O}(1/V)$ corrections, a numerical evaluation of 
$f^\prime_V(x)$, the derivative of $f_V(x)$ respect to $x$.
After a fit of such derivative, an analytical integration is 
performed to reconstruct the desired probability distribution function.
At the very end a multi-precision code is used to compute the sum
(\ref{Z}) and, consequently, all the relevant physical quantities.

The fitting procedure intrinsic in our approach
avoids the pathological flattening behavior of the free energy 
as already noted by other groups \cite{Japan_2}.
It is also important to notice that the multi-precision step needed
to compute the observables
(free energy density and topological charge density) from
the definition (\ref{Z}) is not a bottleneck.
As reported elsewhere \cite{lat02_vic} it can be shown that,
at large volume, only the values of $x$ such that
$f_V(x) \le g_V(\pi)$ (where $g_V(\theta)$ is the exact free energy
density normalized to zero at $\theta=0$) are relevant for the computation
of the sum (\ref{Z}). This has some positive implications:\\ 
$(i)$ 
it is possible to fit only the relevant part of the 
data having very good fits with a
suitably chosen function depending on few parameters;\\ 
$(ii)$ 
the number of terms relevant for the correct evaluation
of the sum (\ref{Z}) is much smaller than $V$ and diminishes 
increasing $\beta$ (recall that $g_V(\theta)$ goes to zero as 
$\beta\to\infty$);\\
$(iii)$
the number of significant digits
necessary for the calculation of (\ref{Z})
is related to the quantity $e^{-Vg_V(\pi)}$
(the smallest value the partition function takes) and, at fixed volume, 
the required precision in the multi-precision step diminishes
when going deeper in the weak coupling region.

To take advantage of it we use an iterative procedure: from a
first ``trial fit'' done
with all the data available we can have an estimation of $f_V(x)$
and, using the multi-precision code, an estimation of $g_V(\pi)$.
Using this approximate information we can extract a trial value $x_{max}$
such that $g_V(\pi) \lesssim f_V(x_{max})$. Having a better determination
of the interval actually necessary for current analysis we can repeat
the procedure steps. 
In practice the convergence is very fast (two, maximum
three steps are enough in all cases) and, at the end, allows to fit 
only the relevant part of the data having very good fits.

The second method \cite{metodo2} uses $x(h)$, the topological charge density, 
in a different way.
It is possible to rewrite the sum (\ref{Z}) as an even polynomial in the
variable $z=\cosh\frac{h}{2}$ and to define the derivative of the free 
energy density respect to $\log z^2$
as $y(z)=x(h)/\tanh\frac{h}{2}$. 
Using the transformation $y_\lambda(z)=y(e^{\frac{\lambda}{2}}z)$
and plotting $y_\lambda/y$ against $y$ one gets typically a smooth function 
for small $y$ \cite{metodo2}. In addition, in the weak coupling scaling 
region, the function $y_\lambda/y$ can be computed from numerical simulations 
at real $h$ until values of $y$ very close to $0$. Hence 
we can expect a simple extrapolation to $y\to 0$ ({\it i.e.} in
the region corresponding to real $\theta$) to be reliable and
thus to obtain $x(\theta)$ 
(of course the result has to be independent of the 
specific value of $\lambda$).

In the same spirit the effective exponent 
$\gamma_\lambda=\frac{2}{\lambda}\log(y_\lambda/y)$ as a function of
$y$ will give the dominant power of $y(z)$ as a function of $z$ near
$z=0$ or, equivalently, the behavior of $x(\theta)$ for $\theta\to\pi$.
If $\gamma_\lambda(y\to 0)=1$ $\textrm{CP}$ symmetry is spontaneously 
broken at $\theta=\pi$, values between 1 and 2 indicate a second
order phase transition with a divergent susceptibility and so on.
Clearly the method is based on the feasibility of a simple extrapolation 
procedure.
This can only be considered a working hypothesis but it has
already been successfully tested in different models \cite{metodo2}.

For both approaches systematic effects can be present: 
${\cal O}(1/V)$ corrections to saddle point solution and the arbitrariness
of the fitting function in the first case (see the relative discussion
in \cite{metodo1}), the reliability of extrapolation
in the second one. What is crucial is that the consistency of results
coming from the two procedures is a strong indication that 
systematic effects are negligible and 
we are actually reconstructing the true
$\theta$ dependence of the model.

For $N=10$ and $h=0$ the scaling window starts at $\beta$ around 0.75 
(see \cite{Campostrini}) and we decided to concentrate our efforts from
this point on. 
The first step is to establish the appropriate lattice sizes.
From $h=0$ simulations at fixed $\beta$ one realizes that the 
asymptotic value of the topological susceptibility is reached when
the lattice side $L$ is roughly larger than $8$ times the correlation
lengths. Using available data \cite{Campostrini} 
we get $L\ge 40$ for $\beta=0.8$,
$L\ge 70$ for $\beta=0.9$ and so on.
Convergence of $\chi_t$ to the infinite volume limit is even faster
at imaginary $\theta$ and consequently we expect the 
results at real $\theta$ extracted from $h\ne 0$ simulations will scale too.

For $h>0$ simulations we considered different
$h$ values ($\simeq 50$ where typically enough) in steps of 0.2.
We are interested in precise measurements of the topological 
charge density as a function of $h$ but, for each configuration, 
this quantity is a discrete number that takes the values $n/V$, where $n$ is 
an integer between 0 and $V$. It turns out that values of $x$ much smaller
than $1/V$ are very difficult to measure and, with affordable
statistics, they suffer huge relative errors. 
The only way to have
reliable measurements of $x(h)$ for the smaller values of $h$ is to
consider large enough volumes. How large can be estimated using a linear
extrapolation around $h=0$: 
if we write $x(h=0.2)\simeq\frac{1}{5}\chi_t(h=0)$ and impose the
condition $x(h=0.2)\ge 1/V$ we can get the minimum $V$ value to be
considered.
Using again the available data \cite{Campostrini} we get that
for $\beta=0.85$,
$L$ has to be at least 100 while for $\beta=0.9$ a $L\simeq200$ lattice
is the best option.

To keep the updating procedure as simple as possible we 
used a standard Metropolis algorithm and the main runs where 
$L=100$, $\beta=0.8,0.85$ with $10^6$ configurations,
$L=200$, $\beta=0.8,0.85,0.9$ with $0.5\, 10^6$ configurations.
\begin{figure}[t!]
\centerline{\includegraphics*[width=2in,angle=90]{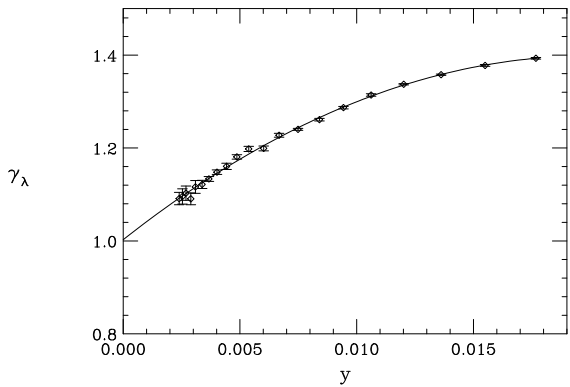}}
\caption{The symbols are $\gamma_\lambda(y)$ in a $L=100$, $\beta=0.80$ lattice
for $\lambda=1.0$. The continuous line is a quadratic fit to the data
(reduced $\chi^2$ is 1.20).}
\label{fig:gamma}
\end{figure}

\begin{figure}[t!]
\centerline{\includegraphics*[width=2in,angle=90]{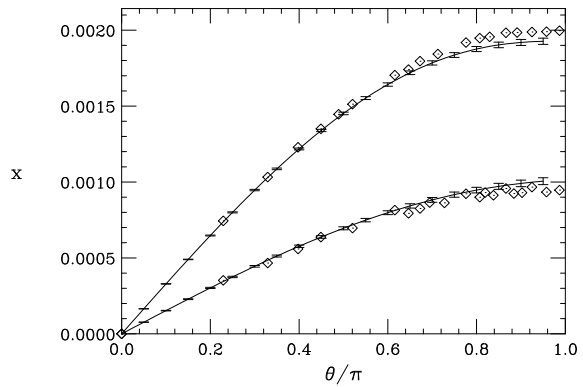}}
\caption{Topological charge density \textit{versus} $\theta/\pi$ at 
$\beta=0.8$, $L=200$ (upper points) and $\beta=0.85$, $L=200$ (lower ones): 
first method (to guide the eyes the continuous curve joins the points) and from
$y_\lambda$ extrapolation (diamonds, statistical errors are smaller 
than the symbol and are not reported).}
\label{fig:confronto}
\end{figure}

In Fig. 1 we present the effective exponent $\gamma_\lambda(y)$ for
the $L=200$ lattice at $\beta=0.80$ and $\lambda=1.0$. 
As for other cases \cite{metodo2} the data points follow a 
smooth curve and a quadratic 
function fits perfectly the data. The extrapolated value for $y\to 0$
is $\gamma_\lambda(0)=1.003(5)$: a sharp indication for a 
spontaneously broken $\textrm{CP}$ symmetry at $\theta=\pi$.
The same holds for different ${\cal O}(1)$ values of $\lambda$ 
and for other lattice sizes and/or couplings.

Following the procedure reported in \cite{metodo2} we can reconstruct,
from the fits of $y_\lambda (y)$, the 
order parameter dependence on $\theta$. The results for the order parameter
at $\beta=0.80$ and $\beta=0.85$, for the $L=200$ lattice, are 
shown in Fig. 2.
Indistinguishable results were obtained using the data of the $L=100$
lattices, as expected if finite volume effects are negligible.
 
On the same plot the output of the first method (based on the
determination of the probability distribution function) \cite{metodo1}
is also 
reported: the agreement between the two procedures is at (few) \% level.
These statements can be extended to the smallest gauge coupling 
available, with the only difference that vertical scale of the plot 
(and consequently the jump of the topological charge 
for $\theta=\pi$) reduces increasing $\beta$.
For both methods the evaluation of statistical errors has been done computing
the physical quantities starting from independent simulation runs.
It comes out that the statistical error on the topological charge 
density grows moderately with $\theta$
and for $\theta\sim\pi$ takes its maximum value that, for the
three values $\beta=$ 0.8, 0.85, 0.90 is of
the order of 1, 2, 5\% for the first method and
2, 3, 5\% for the other one.

The fitting procedure used in both methods are potentially sources
of systematical errors.
Concerning the first method \cite{metodo1}
we remind that, in order 
to extract the probability distribution function, 
we have to fit the $h(x)$ data points. In principle any choice that
respects the known symmetries ($h(x)$ is an odd function) is admissible.
We have tried different function families and the best results were
obtained with $h(x)=(a+bx^2+cx^4)\mathrm{asinh} (dx)$; with this 4-parameter
function we were able to fit the data over the entire $x$ interval we
were interested in, with optimal $\chi^2/d.o.f.$ values.
In particular we got $\chi^2/d.o.f.=0.40$ for $\beta=0.80$,
$\chi^2/d.o.f.=0.62$ for $\beta=0.85$ and
$\chi^2/d.o.f.=0.67$ for $\beta=0.90$.
 
For the second method \cite{metodo2} we also tried several different 
functions for fitting $y_\lambda (y)$, obtaining fits of comparable quality,
and identical results for the order parameter.

Fig. 2 is a {\it a posteriori} check of the validity of our
estimation of errors: since the two approaches are so different each other,
we do not expect agreement between the results if uncontrolled systematics 
plays a major role. This is, in our opinion, the most comfortable 
outcome of the whole procedure.

\begin{figure}[t!]
\centerline{\includegraphics*[width=2in,angle=90]{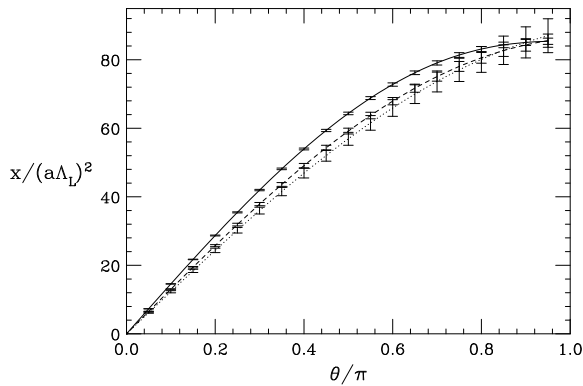}}
\caption{Topological charge density divided by the squared lattice size
(in units of $\Lambda_\textrm{L}$)
\textit{versus} $\theta/\pi$ for $\beta=0.80,0.85,0.90$ 
(continuous, dashed, dotted line respectively) in the $L=200$ lattice. 
}
\label{fig:betas}
\end{figure}

The next question is to check scaling.
In Fig. 3 we plot the continuum topological
charge density, $x(\theta)/a^2$,  
in units of the lattice $\Lambda_\textrm{L}$ renormalization group
parameter, related to the lattice spacing $a$ by the two loop 
perturbative renormalization group equation: 
$a=\Lambda_\textrm{L}^{-1} (2\pi\beta)^{2/N}\exp(-2\pi\beta)$. 
Note that $a^2$ varies by a factor of 4 in the considered
$\beta$ interval.
In Fig. 4 we plot the lattice topological charge density normalized by
the measured topological susceptibility at $\theta=0$.
In both cases we plot the first method results only
since, as stated above, those obtained with the second approach
are practically not distinguishable.

\begin{figure}[t]
\centerline{\includegraphics*[width=2in,angle=90]{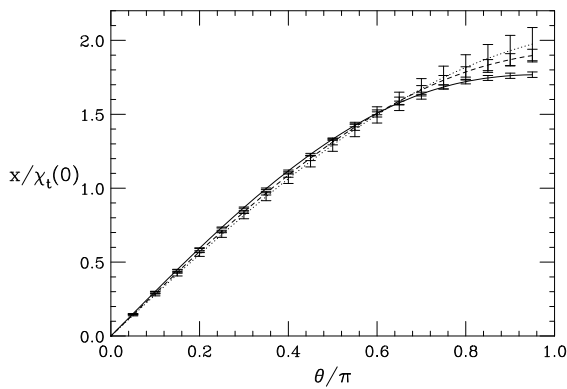}}
\caption{As in Fig. 3 for the topological charge density divided 
by the measured topological susceptibility at $\theta=0$. 
}
\label{fig:beta_np}
\end{figure}

We see a good data collapse of plots at different $\beta$ values.
In Fig. 3 violations to asymptotic scaling are at most $\pm$6\%. 
The origin of these violations could be the higher order corrections
to the perturbative renormalization group equation and then,
as seen in the plot,
we would expect higher violations for the smaller $\beta$ value.
Indeed, as shown in Fig. 4, the non perturbative scaling is 
realized with much better accuracy: the three 
curves are now compatible within statistical errors.
This is again a consistency check of both the results and the method(s).

To conclude we want to emphasize that the continuum $\theta$ dependence of a 
confining and asymptotically free quantum field theory has been
fully reconstructed.  Data collapse for different couplings
within \% level. The evidence for scaling in $\textrm{CP}^9$
model at non zero $\theta$ is the strongest indication that the
$\textrm{CP}$ symmetry is spontaneously broken in the continuum, 
as predicted by the large $N$ expansion.

This work also shows how the inapplicability of the importance sampling 
technique to simulate physically interesting $\theta$-vacuum models can 
be successfully overcome.

The authors thank the Consorzio Ricerca Gran Sasso that has provided 
the computer resources needed for this work.
This work has been partially supported by an INFN-CICyT collaboration and 
MCYT (Spain), grant FPA2000-1252. 
Victor Laliena has been supported by Ministerio de Ciencia y Tecnolog\'{\i}a 
(Spain) under the Ram\'on y Cajal program.

\end{document}